\documentclass[12pt,preprint]{aastex}

\def\au{{\rm au}} 
 
\def\kms{{\rm km}\,{\rm s}^{-1}}

\def\kpc{{\rm kpc}}
\def\mas{{\rm mas}}

\def\pc{{\rm pc}}

\def\rel{{\rm rel}}

\def\e{{\rm E}}

\def\bv{{\bf v}}

\begin{document}
\title{Exploration of M31 via Black-Hole Slingshots and the ``Intergalactic Imperative''}

\author{\textsc{
Andrew Gould$^{1,2}$}}

\affil{$^{1}$Max-Planck-Institute for Astronomy, K\"{o}nigstuhl 17,
69117 Heidelberg, Germany}

\affil{$^{2}$Department of Astronomy, Ohio State University, 140 W.
18th Ave., Columbus, OH 43210, USA}

\begin{abstract}

I show that a gravitational slingshot using a stellar-mass black hole
(BH) orbiting SgrA* could launch robotic spacecraft toward M31 at
$0.1\,c$, a speed that is ultimately limited by the tensile strength
of steel and the BH mass, here conservatively estimated as $m_{\rm
  bh}=5\,M_\odot$.  The BH encounter must be accurate to $\la 1\,$km,
despite the fact that the BH is dark.  Navigation guided by
gravitational microlensing can easily achieve this.  Deceleration into
M31 would rely on a similar engagement (but in reverse) with an
orbiting BH near the M31 center.  Similarly for a return trip, if
necessary.  Colonization of M31 planets on 50 Myr timescales is
therefore feasible provided that reconstruction of humans,
trans-humans, or androids from digital data becomes feasible in the
next few Myr.  The implications for Fermi's Paradox (FP) are
discussed.  FP is restated in a more challenging form.  The
possibility of intergalactic colonization on timescales much shorter
than the age of Earth significantly tightens FP.  It can thereby impact
our approach to astrobiology on few-decade timescales.  I suggest
using a network of tight white-dwarf-binary ``hubs'' as the backbone
of a $0.002\,c$ intra-Galactic transport system, which would enable
complete exploration of the Milky Way (hence full measurement of all
non-zero terms in the Drake equation) on 10 Myr timescales.  Such a
survey would reveal the reality and/or severity of an
``intergalactic imperative''.

\end{abstract}


{\section{{Introduction}
\label{sec:intro}}

At first sight, the problem of exploring and colonizing other
galaxies does not appear to be very pressing.  Even with near-$c$ travel,
it would take 2.5 Myr for a probe to reach the nearest large galaxy
(M31), and another 2.5 Myr to receive any report on what it discovered.
The problems of launching toward (not to speak of then stopping upon arrival 
at) M31 appear to be orders or magnitude more difficult than sending
probes to planets of the nearest stars, which itself is a daunting
enterprise.  Hence, the problem of intergalactic exploration
and colonization seems little more than an intellectual pastime.

Nevertheless, such ``impractical'' intellectual exercises can have
quite practical effects, both in stimulating ideas to achieve
more prosaic aims and in framing our perspective on how to focus
present resources.

Here, I demonstrate a ``practical'' method of launching a space probe
toward M31 at $0.1\,c$ using a black-hole (BH) slingshot.  The same
mechanism could be used to stop the probe in M31 and then relaunch
it back to the Milky Way (MW) after it had carried out its explorations.
The method is ``practical'' in the sense that it does not violate
any laws of physics, relies on our present understanding of materials
science, and does not require exorbitant resources.

{\section{{Slingshot Mechanism}
\label{sec:sling}}

The slingshot mechanism has been used for decades to assist in the
launching of probes into the outer solar system.  For example,
the Pioneer 10 \& 11 and Voyager 1 \& 2 missions (launched 1972--1977)
all used an assist from Jupiter
to reach (and then leave) the outer solar system.  The idea is simple.
By conservation of energy, the satellite must enter and leave the
gravitational influence of Jupiter at the same speed as seen from the
Jupiter frame: only the direction is changed by the encounter with
Jupiter.  However, Jupiter is moving in the solar frame.  If the
angle of the outgoing spacecraft motion is more aligned to Jupiter's
own motion than the angle of the incoming motion, then the velocity
of the spacecraft will increase in the solar frame.

The same principle can be applied to any orbiting body, and indeed
reapplied repeatedly.  For example, NASA was able to launch (in 1989) the very
heavy Galileo probe to Jupiter by making use of four slingshot 
maneuvers, first at Venus, then at Earth, then at Gaspra, then at Earth again.

{\section{{Tidal Limit}
\label{sec:tidal}}

One problem that NASA has not so far encountered in applying the
slingshot method is tidal disruption of the spacecraft.  Tidal
acceleration is basically given by the mean density interior to
the orbit, and the greatest such interior density within the 
solar system would occur in a near passage to Earth\footnote{In principle,
very close passage to a metallic (iron) asteroid could produce
slightly higher tidal forces.}.  
However,
this tidal acceleration would be the same as sitting on Earth.
So, if the spacecraft could be constructed on Earth, then it
would easily survive such tidal stresses.

Nevertheless, for BH slingshots, tidal stress will prove to be
the limiting factor.  Consider a spacecraft in the shape of
a cube of size $\ell$, whose walls have thickness $t\ll\ell$ and composed 
of material of density $\rho$.  When the center of the cube
is at a distance $r$ from a BH of mass $m$, the near and far walls of the
cube will each suffer equal and opposite tidal forces of
$F = (2Gm/r^3)(\ell^2 t\rho)(\ell/2) = (Gm/r^3)\rho\ell^3 t$.
This tidal force will have to be supported by walls with total area
$A = 4\ell t$.  Hence, the two faces alone require support of
$F/A = (Gm/r^3)\rho\ell^2/4$.  The four walls themselves must also be
supported.  These have two times larger mass but only 1/3 the mean
tidal acceleration.  Hence, $F/A = (5/12)(Gm/r^3)\rho\ell^2$.  Finally,
the spacecraft should have some content (not just walls).  I
parameterize this by $F/A = \zeta(Gm/r^3)\rho\ell^2$, where $\zeta$
is of order 1 or 2.  I evaluate this in a way that can be
related to material properties,
\begin{equation}
{F/A\over\rho} = \zeta{Gm\over r^3}\ell^2
= \zeta{Gm\over r_g}\biggl({\ell\over r_g}\biggr)^2{1\over n^3} 
= \zeta\biggl({\ell\over r_g}\biggr)^2{c^2\over n^3} 
\label{eqn:stress}
\end{equation}
where $r_g$ is the gravitational radius of the BH, and $r = n r_g$
is the spacecraft's distance from the BH.

We can now equate this to
\begin{equation}
V_{\rm steel}^2 = {T_{\rm steel}\over \rho_{\rm steel}} = 
{2.53\,{\rm GPa}\over 8\,{\rm gm/cm^3}} = (0.56 \,\kms)^2 =
3.5\times 10^{-12} c^2 
\label{eqn:vsq}
\end{equation}
where $T_{\rm steel}$ is the tensile strength of steel, $\rho_{\rm steel}$ is
its density and GPa$=10^9\,$newtons/m$^2$. Equating $T$ with $F/A$
yields
\begin{equation}
n = \biggl(\sqrt{\zeta}\,{\ell\over r_g}\,{c\over V_{\rm steel}}\biggr)^{2/3}
  = 80\,\zeta^{1/3}\biggl({r_g/\ell\over 750}\biggr)^{-2/3},
\label{eqn:neval}
\end{equation}
where I have normalized the evaluation to an $\ell=10\,$m box
passing by an $m=5\,M_\odot$ ($r_g=7.5\,$km) BH.  We see that the
result depends only weakly on $\zeta$.  Before, proceeding, I
note that $V_{\rm steel}$ is well below the speed of sound in steel,
which is roughly $3\,\kms$.

{\section{{Black-hole Slingshot Mechanism}
\label{sec:bh_sling}}

I consider a small black hole (BH) of mass $m$ in a circular
orbit at $N$ gravitational radii of a larger BH of mass $M\gg m$, where
$N\gg 1$, i.e., at velocity $v=c/\sqrt{N}$.

I now calculate the orbit of a test particle that ``falls'' from
infinity, and passes the smaller BH at peribothron $q_{\rm bh}$ of $n$
gravitational radii (of the smaller BH).  However, rather
than directly ``falling'', i.e., on a radial orbit, I will
consider the more general case that the test particle has
some angular momentum so that its trajectory intersects
that of the small BH at an arbitrary angle $\phi$ as seen
in the frame of the large BH.

Normally, the Rutherford scattering angle is written,
\begin{equation}
\label{eqn:rutherford}
\psi = 2\tan^{-1} {Gm\over v_0^2 b},
\end{equation}
where $b$ is the impact parameter and $v_0$ is the velocity
at ``infinity'' (i.e., away from the potential of the central mass).
However, in our case, the controlling variables are $v_0$ and $q_{\rm bh}$
(the peribothron),
the first being set by the geometry of the problem and the second
by the tidal constraints from Section~\ref{sec:tidal}.  By imposing
energy and momentum conservation, we obtain $b=q\sqrt{1+2/\xi}$, and so
Equation~(\ref{eqn:rutherford}) becomes
\begin{equation}
\label{eqn:rutherfordx}
\psi = 2\cot^{-1} \biggl(\xi \sqrt{1 + {2\over\xi}}\biggr)
     = 2\csc^{-1}(1 + \xi);
\qquad \xi \equiv {v_0^2 q_{\rm bh}\over Gm} \rightarrow n{v_0^2\over c^2},
\end{equation}
or
\begin{equation}
\cos\psi = 1-{2\over (1+\xi)^2};
\qquad
\sin\psi \rightarrow {2\over 1+\xi} - {1\over (1+\xi)^3} - {1\over 4(1+\xi)^5}
\ldots
\label{eqn:rutherford2}
\end{equation}

In the frame of the small BH, the test particle will have
incoming velocity\break
$\bv_0 =(1-\sqrt{2}\cos\phi,\sqrt{2}\sin\phi)c/\sqrt{N}$,
i.e., $v_0/c= \eta/\sqrt{N}$, and at angle $\delta$ defined by\break
$\cos\delta = (\sqrt{2}\cos\phi - 1)/\eta$,
$\sin\delta = \sqrt{2}\sin\phi/\eta$, with 
$\eta^2\equiv (3 - \sqrt{8}\cos\phi)$.  Hence, $\xi = \eta^2 n/N$.
Then, by the law of cosines, the exit speed squared (in the large-BH frame) is,
\begin{equation}
{v_1^2\over c^2} 
= {\eta^2 +1 \over N} -2{\eta\over N}\cos(\psi + \delta),
\label{eqn:v1}
\end{equation}
and hence it will escape to infinity at speed 
\begin{equation}
{v_2^2\over c^2} 
= {\eta^2 -1\over N} -2{\eta\over N}\cos(\psi + \delta) 
= {1\over N}\biggl(2-\sqrt{8}\cos\phi - 2\biggl(1 - {2\over (1+\xi)^2}\biggr)
(1-\sqrt{2}\cos\phi) + \sqrt{8}\sin\phi\sin\psi\biggr)
\label{eqn:v2}
\end{equation}
\begin{equation}
= {1\over N}\biggl({4(1-\sqrt{2}\cos\phi)\over (1+\xi)^2} + 
\sqrt{8}\sin\phi\sin\psi \biggr).
\label{eqn:vp}
\end{equation}
Figure~\ref{fig:v2} illustrates Equation~(\ref{eqn:vp}) for various
values of $N/n$.

To gain analytic insight, I now consider the special case of $\phi=90^\circ$,
i.e., that the test particle is initially falling directly into the
large-BH potential.  Then, $\xi\rightarrow 3n/N$, and
Equation~(\ref{eqn:vp}) becomes,
\begin{equation}
{v_2^2\over c^2} = {4/(1+\xi)^2 + \sqrt{8}\sin\psi\over N} =
 {\sqrt{32}\over N + 3n} + {4N\over (N + 3n)^2} - {\sqrt{8} N^2\over (N+3n)^3}
+ {\cal{O}}\biggl({N\over N+3n}\biggr)^5 .
\label{eqn:vp2}
\end{equation}
The derivative of the first two terms with respect to $N$ is 
strictly negative, and I have confirmed that the
exact expression falls monotonically with $N$ for $n\gg 1$.
Hence, for fixed $n$, the maximum ejection speed is approached
for any value $N<n$.  For example, for $N=n$
$(\xi =3,\ \sin\psi=0.176)$,
the ejection speed is already 86\% of the maximum possible,
\begin{equation}
v_{2,\rm max} = \biggl({32\over 9}\biggr)^{1/4}{c\over\sqrt{n}},
\label{eqn:v2max}
\end{equation}
where $n$ is the closest allowed peribothron, which is set by the physical
properties of the spacecraft and the mass $m$ of the smaller BH.

Combining this with the estimate $n\sim 100$ from Section~\ref{sec:tidal},
we obtain $v_{\rm final}\sim [(32/9)^{1/4}/\sqrt{100}]c = 0.14\,c$.
More conservatively, I adopt $v_{\rm final} \sim 0.1\,c$ as feasible.

Of course, it would not be possible to launch toward M31 
at Galactic coordinates $(l,b)=(+121.2,-21.5)$ via slingshot from
a small BH in a circular orbit from a strictly radial spacecraft
orbit.  However, one could roughly reverse
the geometry simply by sending the spacecraft into a nearby encounter
with the supermassive BH and arranging the encounter with the small BH
on the return.
In the extreme limit, this would reduce the required deflection from
$120^\circ$ to $60^\circ$.  With a less extreme approach to the supermassive
BH, the required deflection angle could be further reduced.

I do not pursue this issue in detail because the real population of
small BHs will not be on circular orbits, and moreover, the parameters
of these orbits (at the time of the encounter) will be completely
unknown at the time the mission takes off from the solar system.
The real problem will therefore be precision mapping of the orbits
(and masses) of this population and then choosing the best small BH for
the slingshot.  I now turn to this problem.

{\section{{Mapping the BH Population at the Galactic Center}
\label{sec:map}}

BHs do not emit radiation and so are, per se, ``invisible''.
While in principle they do emit \citet{hawking75} radiation, the 
temperature of an $M\sim 5\,M_\odot$ BH is $\sim 10^{-8}$ smaller
than that of the cosmic microwave background (CMB).  Hence, what could
be seen, in principle, is a ``hole'' in the CMB.  However, resolving
this hole from an observatory in the solar system
would require an apparatus with dimension 
$D\sim (\lambda/r_g)R_0 \sim 200\,\au$, with sufficient collecting
area to make an image during the BH $r_g$ self-crossing time, roughly
1 ms.  Here, $\lambda$ is the radiation wavelength and 
$R_0$ is the Galactocentric distance.  One could imagine looking
for the ``hole'' in $K$-band light from some background source
(such as SgrA*), which would reduce the dimensions of the apparatus
by a factor 500, but would still require a huge collecting area for
the 1 ms imaging against high-background.

Without placing doubts on the capabilities of our descendants,
I think that the prospects for such projects are remote.  However,
from the current standpoint, the main issue is not their practicality,
but their utility for planning a slingshot mission.  The BH orbital
parameters are not needed at the time of launch from Earth, but rather several
Myr later at the time of the slingshot boost.  Of order a billion BH orbits will
elapse during this interval.  As we will see, the 3-space position
of the BH must be predicted to of order 1 km.  Hence, even a thorough
mapping of the Galactic center from Earth would be of little use
in preparing a slingshot.  On the other hand, using the technique
of gravitational microlensing \citep{einstein36}, it is quite
feasible to obtain a detailed map of the BH distribution as the
spacecraft approaches the Galactic center\footnote{J.\ Carpenter
was the first to suggest that gravitational lensing could
play a key role in travel to M31, but he did not present any
details (``They Live'', 1988).}.  In this context,
constraints on the overall small-BH population that is orbiting
SgrA* would be important to planning the mission, and this could
be achieved with a combination of microlensing and gravitational
wave observations.  However, the details of such an investigation
lie beyond the scope of the present work.

{\subsection{{Microlensing Basics}
\label{sec:basics}}

For present purposes, the main required microlensing concepts are the
angular Einstein radius, $\theta_\e$, and the Einstein timescale, $t_\e$,
\begin{equation}
\theta_\e =\sqrt{\kappa M_L \pi_\rel};
\qquad
t_\e = {\theta_\e\over \mu_\rel};
\qquad
\kappa = {4G\over c^2\au}\simeq 8.14\,{\mas\over M_\odot},
\label{eqn:thetae}
\end{equation}
where $M_L$ is the lens mass and ($\pi_\rel,\mu_\rel$) are the
lens-source relative (parallax, proper motion).  If the lens-source
angular separation (normalized to $\theta_\e$) is $u$, then the lens
will split the source light into 
two images, one outside the ``Einstein ring'' on the
same side as the source, and one inside the Einstein ring,
on the opposite side, with
positions and magnifications
\begin{equation}
\theta_\pm = \theta_L + {u\pm\sqrt{u^2 +4}\over 2}\theta_\e;
\qquad A_\pm = {A\pm 1\over 2};
\qquad A = {u^2+2\over u\sqrt{u^2 + 4}}
\label{eqn:thetapm}
\end{equation}
The key points about these equations are first, that $A\rightarrow 1/u$
and $A\rightarrow 1 + 2/u^4$ for $u\ll 1$ and $u\gg 1$, respectively.
Hence, the magnification can be significant only when the source is inside
or near the Einstein ring.  And second, for the first limit
the images are of about equal brightness and both are close to the Einstein
ring, while in the second limit, the major image is close to the source
position, while the minor image is highly demagnified, $A_- \rightarrow u^{-4}$.

{\subsection{{BH Mapping from Earth: Microlensing Blind spot}
\label{sec:blind}}

The small-BH population that very likely orbits SgrA* \citep{gme00}
cannot be mapped by microlensing observations from Earth.  There
is a general problem that the probability that any given BH is magnifying
a background star is very low, so that successive microlensing events,
likely separated by thousands of years, cannot be distinguished from
other events due to other BHs.  But there is also a more fundamental
problem.  For a source star with relative lens-source distance
$D_{LS} = D_S - D_L \ll D_L$, the physical Einstein radius $r_\e = D_L\theta_\e$
due to SgrA* itself is well approximated by,
\begin{equation}
r_{\e,\rm SgrA*}\simeq 180\,\au \sqrt{D_{LS}\over 1\,\pc}
= 4500\,r_{g,\rm SgrA*}\sqrt{D_{LS}\over 1\,\pc}.
\label{eqn:resgra*}
\end{equation}
Imaging by additional lenses (such as the small BHs that orbit SgrA*)
that lie well outside this radius are not qualitatively affected by SgrA*
lensing.  For small BHs that are projected exactly on, or very close
to $r_{\e,\rm SgrA*}$, detection is quite feasible and has
interesting features.  Mathematically,
this is the regime of resonant caustics in ``planetary microlensing''. 
I discuss this regime in Section~\ref{sec:planet}.  However, small BHs
that lie well within this radius very rarely 
leave any detectable microlensing signature.
Thus, the small BHs that are detectable from Earth, with orbital
velocities $v/c <(4500)^{-1/2}\sim 0.015$ are not
relevant to the slingshot mechanism.  That is, such boosts are only
halfway (in log) between what can actually be achieved and what could
be achieved by slingshots from local white-dwarf (WD) binaries,
without going to the Galactic center, as I discuss in Section~\ref{sec:sketch}.

{\subsection{{In-Flight BH Mapping}
\label{sec:flight}}

The blind spot 
will only begin to substantially contract relative to 
Equation~(\ref{eqn:resgra*})
when the spacecraft comes within 1 pc of SgrA*.  At this point, we
can also begin to consider the general populations of bulge sources,
which are both much more distant, $D_S\sim 1\,\kpc$, and much more numerous
than those in the nuclear star cluster.  In this regime, the formula
is similar to Equation~(\ref{eqn:resgra*}), but with $D_{LS}\rightarrow D_L$.
\begin{equation}
r_{\e\rm SgrA*}\sim 180\,\au \sqrt{D_{L}\over 1\,\pc}
= 4500\,r_{g,\rm SgrA*}\sqrt{D_{L}\over 1\,\pc}.
\label{eqn:resgra*2}
\end{equation}
Hence, the innermost relevant BHs will begin to ``come into focus''
at about $D_L\sim 100\,\au$, which corresponds to two weeks prior to
slingshot (when the infalling speed is already $0.03\,c$).

I analyze the observations required to measure the BH masses and
orbits within the framework of isolated small-BH lenses.  In fact,
when the small-BH first comes into focus, it will act like a
``microlensing planet'' from the standpoint of the microlensing
formalism.  That is, the positions and magnifications of the 
microlensed images can only be understood as a product of the
combined influence of the small-BH ``planet'' and its SgrA* ``host''.
Their mass ratio, $q\sim 10^{-6}$ happens to be just below that of the most
extreme current detections \citep{kb180029,ob190960,kb200414,kb191053}.
Nevertheless, from the standpoint of understanding the basic
principles, it is best to begin with the simpler case.  I will
then take account of ``planetary'' microlensing in Section~\ref{sec:planet}.
Moreover, the case of isolated small-BH microlensing becomes
increasingly relevant as the spacecraft approaches slingshot.

At these epochs (when $D_L < 100\,\au$), the small-BH angular
Einstein radius and Einstein timescale would be
\begin{equation}
\theta_\e = 9^{\prime\prime}\biggl({D_{L}\over 100\,\au}\biggr)^{-1/2};
\qquad
t_\e = 20\,{\rm s}\biggl({D_{L}\over 100\,\au}\biggr)^{1/2};
\label{eqn:resmall}
\end{equation}
where I have assumed that the small BH is moving at $0.1\,c$ across the
line of sight.  The main population of sources will be
the Galactic bulge dwarf stars, with a characteristic distance
of $D_S\sim 1\kpc$, and with a surface density for a very 
conservatively chosen $M_K<4$ (i.e., 
$K_0<14$, and so $K<16$ allowing for $A_K=2$) 
of $N_{K<16}\sim 0.05\,{\rm arcsec}^{-2}$.  Hence, at any given
moment, there would be of order 12 stars inside the Einstein ring,
including of order 3 that were magnified by at least a factor 2 (so
with counter-image $A_- = (A-1)/2\geq 0.5$, at least half as 
bright as the source).
The main potential problem is that the lens is moving relative to the
source at about $2^{\prime\prime}/{\rm s}$ (and the magnified images
can be moving even faster), so that exposure times should be kept
to e.g., $0.1\,$s.  However, it is not necessary that the individual
images have high signal-to-noise ratio (SNR) to reconstruct the lensed-image
brightness and trajectory from a high-cadence series of observations.
Thus, effectively, the integrated SNR would be that of a $20\,$s exposure
of a $K\sim 16$ star, which would yield very precise results even 
with a small (e.g., 2m) telescope.

As noted above, the ``final descent'' toward SgrA* would take
about 2 weeks.  During this time, there would be
a near continuous series of such measurements, with repeated, but
ever shorter interruptions as the lens passed through the SgrA* blind spot.
These observations would cover about 10 orbits and would comprise
about about $6\times 10^4$ individual epochs.

I now examine what is learned from each observation and how the
information from these observations can be combined to determine
the BH orbit to high precision.  I will argue below that the required
precision is $\sim 1\,$km, including, in particular, the distance
between the spacecraft and the BH along the line of sight.

Each ``observation'' is actually a track of observations as the images
streak across the Einstein ring.  Nevertheless, from the standpoint
of understanding the basic physics, I treat these observations
as though they were a single observation at the impact parameter
$u_0$.  In fact, there will be somewhat more information than
is accounted for in this simplified presentation.

The first point is that each observation, by itself, gives a measurement
of $\theta_\e$ from the offset of the two images.  
From Equation~(\ref{eqn:thetapm}),
\begin{equation}
\theta_\e = {\theta_+ -\theta_-\over \sqrt{4+u_0^2}} = 
{\theta_+ -\theta_-\over \sqrt{2 + 2/\sqrt{1-1/A^2}}} .
\label{eqn:thetaedif}
\end{equation}
The total magnification at peak (or equivalently $u_0$)
can be determined, as usual, from fitting the microlensing event.
However, in contrast to current microlensing experiments, the
probability of there being blended light within the point-spread function
(PSF) that does not participate in the event is very low because the
Einstein radius is much larger than the PSF.  In any case, the magnification
can be determined from the flux ratio $r = F_-/F_+$ of the two images, 
$A = (1+r)/(1-r)$.  And the three methods can be cross checked
for consistency.

The lens position can be derived from the mean position of the two images.
\begin{equation}
\theta_L = {\theta_+ +\theta_-\over 2} - u_0\theta_\e,
\label{eqn:thetaesum}
\end{equation}
where $\theta_\e$ can be determined from Equation~(\ref{eqn:thetaedif})
and, again, $u_0$ can be derived from $A$ or $r=F_-/F_+$.  Then,
adopting a total SNR = 100 and $0.25^{\prime\prime}$ FWHM,
each measurement would yield a measurement of $\theta_\e$ with fractional
precision $10^{-5}$ and a lens angular-position measurement with precision
of order 1 mas.  The latter corresponds to about 75 km from 100 AU.
While this is substantially larger than 1 km, the combined information
from $3\times 10^5$ measurements ($6\times 10^4$ epochs, each with $\sim 5$
background stars), constrained by the Kepler equations of the
small-BH orbit, would satisfy this requirement.
Note that a series of 2-dimensional position measurements alone (combined
with the orbital equation) are sufficient to solve for the orbital parameters
(up to a sign flip in the orientation of the orbit).

The interpretation of these measurements requires first that they be
placed on an absolute angular coordinate system, and second that the
distance $D_L$ be known at high precision (to convert angular measurements
made at many different distances into a common physical system).

The first is easily achieved via the Einstein ring of SgrA*.  This
will be delineated with excellent precision in every image by the tens
of thousands of highly magnified $A>100$ images lying within 
$10^{\prime\prime}$ of the $2.5^\circ$ Einstein radius of SgrA*.  These
include also the much more numerous $M_K\la 8$ sources that are
magnified into view.  Hence both the position and Einstein radius
of SgrA* can be determined in the same way as that of the small BH,
but more precisely.

Strictly speaking the Einstein radius is a function not only of the
mass and distance to the lens, but also the distance to the source,
$\theta_\e^2 = \kappa M\pi_\rel \propto (M/D_L)(1 - D_L/D_S)$.  However,
the last term is extremely small, 
$D_L/D_S \sim 100\,\au/1\,\kpc\sim 10^{-6}$ and, in addition, the
individual distances to sources would be known quite well from
millenia of astrometric observations of the field as the spacecraft
approached SgrA*.  So, for purposes of discussion, I just replace
$\pi_\rel\rightarrow \pi_L = \au/D_L$.

Then, for any given observation, the BH mass ratio $q=m/M$ is given by
\begin{equation}
q = {\theta_{\e,\rm small-BH}^2\over\theta_{\e,\rm SgrA*}^2}\,
{D_{\rm SgrA*}\over D_{\rm small-BH}} .
\label{eqn:qbh}
\end{equation}
Note that while, in the context of Earth-bound planetary microlensing,
the distance ratio in the last term is unity to high precision, it can
differ significantly in the present case.  This is due both to the fact
that the small BH is moving closer and farther in its orbit and also
because they are viewed at different angles.  Nevertheless, this
ratio is known from the overall geometry of the system.
As the mass of SgrA* is already known today to 0.3\% precision,
$M_{\rm SgrA*}= 4.154\pm 0.014 \times 10^6\,M_\odot$
\citep{gravitymass}, and this is still
improving using current instruments, the small-BH mass would also be
known  to extremely high precision.
Moreover, each image, with its concomitant measurement of the
angular Einstein radius of SgrA*, will give a precise determination
of $D_{\rm SgrA*}$.  

As mentioned above, the orbit of the small BH can be solved from a series
of 2-D measurements (up to a sign flip in the plane of the sky).  This
degeneracy is easily resolved because $\theta_{\e,\rm small-BH}$ will be
larger when the small BH is on the near side of the plane defined by SgrA*.

{\subsection{{Effects of ``Planetary Microlensing'', etc.}
\label{sec:planet}}

In fact, there will be many effects that are not captured by the
simple approximations of the previous section.  I list the principal
ones here.  However, the main point is that they are all deterministic
and well understood.  Hence, while they require more complex calculations than
those described above, they do not in any way modify the
feasibility of the approach.

Overall, the main modification is that the observations will almost
all be in the regime of ``wide planetary microlensing'',
i.e., $s>1$, where $s$ is the separation between SgrA* and the small BH
in units of Einstein radius of SgrA*.  (For a review of planetary microlensing,
see \citealt{gaudi12}.)\ \ For example, the magnification pattern generated
by the small BH, will not be according to the simple, axisymmetric
formulae of Equation~(\ref{eqn:thetapm}).  Rather, in the presence of
the larger-mass ``host'', it is characterized by a quadrilateral
caustic structure of width (in units of the SgrA* Einstein radius) 
$w\sim 2q^{1/2}/s^2$ (see Equations~(6) and (7) of \citealt{han06}
for more precise estimates).  This corresponds to $w_p = w/q^{1/2}=2/s^2$
in units of the small-BH Einstein radius (where $s$ is still in units of
the large-BH Einstein radius).

Hence, the formalism of the previous
section would only approximately apply at $s\ga 4$.
For example, for $s=2$,  $w_p =0.5$.  So, for the simple
well-magnified ($u_0=0.5$) example described in the previous section,
the source magnification and position would be significantly affected
by the four ``magnification ridgelines'' extending from the cusps of the
caustic.  And, for about half of such events (i.e., $u_0<0.25$), the
two images would suddenly be joined by two additional images as the
source crossed the caustic.  These would appear as a highly 
magnified pairs on opposite sides of the ``critical curve'' 
(the mapping of the caustic onto the image plane) whenever the source 
was just inside the caustic, and they would disappear when the source
again crossed to go outside the caustic.  
Hence, the mathematics of determining the lens position
would be very different from that outlined in the previous section,
but would still be unambiguous.

Another effect, which is generally not faced in planetary microlensing is 
that ``host'' and ``planet'' could not be safely approximated as lying in the
plane of the sky because the difference in their distances might be
of order 1\% or more.  Nevertheless, it is straightforward
to take account of this using standard ray-tracing techniques.  Finally,
microlensing calculations almost always assume non-relativistic motion.
In fact, at the $\sim 0.1\,c$ velocities being considered here, this
approximation is quite good, but probably not good enough for the
$1\,$km (i.e., 0.1\%) precision envisaged.  Nevertheless, even
the fully relativistic problem is well-understood today \citep{zheng00}.

{\section{{Shepherd Spacecraft}
\label{sec:shepherd}}

While the microlensed image observations described in 
Section~\ref{sec:flight} would identify all BHs in the neighborhood
of SgrA*, and also yield very precise measurements of their masses
and orbits, they might not take place soon enough to enable the
spacecraft to make in-flight maneuvers to be properly ``launched''
toward M31 by BH slingshot.  That is, the observations only become possible
very late, when the small-BHs move out of the SgrA* blind spot.  See
Sections~\ref{sec:blind} and \ref{sec:flight}.

This problem can easily be solved by having a second (``shepherd'')
spacecraft, which serves to guide the primary spacecraft toward
its launch site.  The shepherd spacecraft would travel a few hundred
au in front.  It would make a map of all the small BHs and choose
the best one for the launch.  Then, it would radio this information
back to the primary spacecraft.

After carrying out this mission, the shepherd spacecraft could
plunge deep into the potential well of SgrA*, which would enable
it to head back toward Earth.  On its return, it could receive a
report from the primary spacecraft summarizing the success
(or otherwise) of its launch.  Then it could convey this message
to Earth when it reached the solar neighborhood, a few Myr later.

{\section{{Sketch of a Journey}
\label{sec:sketch}}

As far as we are currently aware, the closest supermassive BH
is SgrA*, at the Galactic center, roughly 8 kpc from Earth.  Assuming that
the spacecraft were constructed on or near Earth, it would
have to first be transported to the Galactic center before being
launched toward M31.  Because M31 is 100 times farther than SgrA*,
the spacecraft should not travel much slower than 3\% of the speed of
its intergalactic voyage, i.e., $v_{\rm local}\ga 0.003\,c$,
so as not to contribute substantially to the total time.  It is
quite possible that during the several thousand years before such
a spacecraft is constructed, the means will be developed to achieve
such launch speeds from Earth (or its environs).  However, here
I point out that a sling-shot boost from a white-dwarf (WD) binary
could yield a velocity of this order.  For example, a binary
composed of two $0.6\,M_\odot$ WDs of radius $R_{\rm WD}\sim 10,000\,$km
and separated by $a=j\,R_{\rm WD}$ would have an internal velocity
(relative to the center of mass) of $v=2000\,j^{-1/2}\,\kms$.

The merger time from gravitational-wave emission for such a system is,
\begin{equation}
\Delta t_{\rm merge} = {5\over 512}\biggl({a c^2\over G M_{\rm WD}}\biggr)^3
{a\over c} \sim 0.15\biggl({j\over 10}\biggr)^4\,{\rm Myr}.
\label{eqn:tmerge}
\end{equation}
The travel time from Earth to this ``booster system'' can be somewhat
longer than the merger time, but not dramatically longer;
otherwise small errors in the prediction of its evolution would be
catastrophic.  These times would be equal for $j=10$, a booster-system
distance of 30 pc, and an initial velocity of $200\,\kms$. The separation
$j=10$ corresponds to internal velocities of $v\sim 600\,\kms=0.002\,c$.
Note that the tidal
forces experienced in such an encounter would be substantially smaller
than those of its final launch from the Galactic center.

Navigation would require some fuel, particularly for fine steering
toward the encounter with the local WD binary, and then toward the small-BH
in preparation for slingshot launch toward M31.  However, overall
fuel consumption could be minimized by utilizing two effects.  First,
very small changes in the angle of the trajectory could be made
by deflection off the interstellar medium.  Second, in the dense
environment of the Galactic center, major deflections would be possible
from gravitational encounters with BHs (before or after encountering
the main slingshot BH).  I have calibrated my calculations
to the requirement of 1 km ($\sim 0.1\,$) precision encounter with
the small BH, which could potentially lead to a $10^{-3}$ radian
error in the angle of approach to M31.  It seems plausible that
such small corrections could be achieved in the dense environment 
of the Galactic center.  However, if not, these calculations showed
that the SNR of the microlensing measurements would permit
encounters that are an order of magnitude more precise.

Once arriving at M31, it would be necessary to slow down in order
to explore that galaxy.  This could be achieved by applying the
inverse process to the small BHs orbiting the 
$M_{\rm M31-BH}\sim 3\times 10^7 M_\odot$
supermassive BH at M31's center.  The problem of navigating the
approach to this ``anti-slingshot'' event would be 
more challenging than the MW slingshot in the sense that the
approach speed (at the time that the blind spot became small enough
to map the small-BH distribution)
would be $0.1\,c$, rather than $0.03\,c$.  However,
these difficulties would be significantly ameliorated by the fact
that the M31-center BH is larger by a factor $M_{\rm M31-BH}/M_{\rm SgrA*}\sim 7$.
Hence, for fixed orbital velocity of a small BH, its semi-major axis
would be 7 times larger, meaning that this orbit would start
``coming into focus'' at a distance that is $\sqrt{7}\sim 2.6$ times larger than
in the case of the MW.  Thus, these two factors roughly cancel, meaning
that the overall level of difficulty will be similar.
Thus, it will be essential for the
spacecraft to launch a daughter probe that would map the M31 BH
system in advance of its own arrival, similar to the ``shepherd''
spacecraft described in Section~\ref{sec:shepherd} for the encounter
near Sgr A*.

After slowing to a reasonable velocity, several courses of action
could be pursued, depending on the level of technological development.
At the low-end, the spacecraft could navigate around M31 using WD
binaries, and make flybys of planetary systems that had been identified
from Earth.  At a higher end, it could land on a planet and then use
local materials plus stored DNA sequences, possibly mixed with computer/robot
designs to begin colonizing that planet.  After this colony was
established, it could serve as the center of a colonization system
of many M31 planets.  These colonies might be able to send
electromagnetic signals back to Earth.  But even if not, they
could launch a similar slingshot back to Earth using the M31-center BH.
In this case, it would be unnecessary to go through SgrA*, because the
information learned from M31 could be communicated in a $0.1\,c$ Earth
flyby.  Similarly, for the case that the spacecraft never landed, but
simply explored, it could return toward Earth via an M31-center slingshot.

Assuming $0.1\,c$ intra Local-Group travel, the whole mission would
take about 70 Myr, plus an optional 25 Myr to return to Earth.
See Table~\ref{tab:tab1}.

{\section{{Proof of Principle and Competing Approaches}
\label{sec:proof}}

The main point of this paper is to demonstrate that travel
by robotic spacecraft to M31 on timescales that are short compared
to the age of Earth (though long compared to the time
of human divergence from apes) is feasible.  Further, colonization by
human, trans-human, or android beings is plausibly feasible,
although this would rest on still-to-be-developed technology
of digital-based reconstruction of such beings.  Before
addressing the implications of these feasibility arguments, I
first acknowledge that M31 exploration may well be carried out using
some other technology.  I present one other example here, including
both its advantages and drawbacks.  However, this is really just
for illustration.  Whether slingshots are used, or some other
approach, the conclusion remains that intergalactic exploration
is feasible on 70 Myr timescales, or perhaps shorter.

Another method would be to launch (somehow, see below) a spacecraft
directly toward M31 that accelerated for some time $\Delta t$ at, e.g.,
$1\,g$, and then decelerated for a similar duration as it approached
arrival.  I parameterize $\beta = \tanh\Theta$, 
where $\beta = v/c$ and $\gamma = (1-\beta^2)^{-1/2}$.
During the acceleration phase, and using the time coordinate
$t$ of the accelerated frame, the spacecraft would achieve
$\beta\gamma = \sinh\Theta$, with $\Theta =gt/c$.  I then
find that it would
have traveled a rest-frame distance $d$ in accelerated time $t$ of,
\begin{equation}
d(t) = \int_0^t dt^\prime\,c\sinh\Theta^\prime = {c^2\over g}(\cosh\Theta-1),
\label{eqn:doft}
\end{equation}
or, equivalently,
\begin{equation}
\hat t = \cosh^{-1}(\hat d + 1) \longrightarrow \ln(2\hat d);
\qquad
\hat t\equiv {t\over c/g};
\qquad
\hat d\equiv {d\over c^2/g},
\label{eqn:tofd}
\end{equation}
where the second form applies to the limit $\hat d\gg 1$.
Note that, because $c/g=0.97\,$yr, $\hat t$ can be thought of as time
in years and $\hat d$ can be thought of as distance in light-years.
Thus, to accelerate half-way to M31 and then decelerate for the remaining
half of the trip, would require a total time of 
$2\Delta t \simeq 2(c/g)\ln({\hat d}_{\rm M31}) \sim 29\,$yr, i.e., well within
a (current) human lifetime.  The main, though not only, difficulty
with this approach is the prodigious fuel requirements.  Assuming
a perfect engine, i.e., one that converted mass directly to photons
(or highly relativistic particles), which were then ejected in the
direction opposite to acceleration, the mass of the spacecraft
would evolve as $d M/dt = -(g/c) M$, so that after the first half
of the voyage
\begin{equation}
M_{\rm init} = M_{\rm 1/2}\exp(\Delta\hat t)\longrightarrow M_{1/2}\hat d .
\label{eqn:moft}
\end{equation}
Then, similarly, for the deceleration, $M_{1/2} = M_{\rm final}\hat d$.
Hence, $M_{\rm final} = (\hat d)^{-2} M_{\rm init}$, i.e., a factor $7\times 10^{12}$
lighter.  Other problems (besides inventing the engine) include keeping
teratons of fuel (likely antimatter) stable during the 30-year voyage, and
dealing with the light from M31, which would be Doppler-boosted to the MeV
range.  While we cannot rule out that such challenges will be met
by our descendants, neither can we base our assessment of future
exploration prospects on their success.

However, if the first step could be achieved, i.e., inventing the engine.
then with just 0.1 year acceleration and deceleration phases, the
$0.1\,c$ velocities that are accessible to BH slingshots, would be
achieved.  Using a ``perfect engine'', the fuel requirement would
be only about 20\% of total mass.  The main remaining problem would
be keeping the anti-matter stable for 25 Myr.

Moreover, if this could be achieved, it would be a relatively small
step to increase the acceleration and deceleration phases to
$\Delta t=1\,$yr.  This would yield a coasting speed of 
$\beta = 1/\sqrt{2}$, which would reduce the travel time (seen from
Earth) to $\sqrt{2}d_{\rm M31}/c = 3.7\,$Myr, and the fuel storage time
to 2.6~Myr. The initial-to-final mass ratio would be a very reasonable
$M_{\rm init}/M_{\rm final}=e^2=7.4$.

The point is that while BH slingshots may not ultimately be the method
of choice for intergalactic travel, they show that such travel is feasible.
The feasibility of such travel has both near-term and long-term implications.

{\section{{Fermi's Paradox Revisited}
\label{sec:fermi}}

Fermi's Paradox (FP) is said to have originally been formulated in
just three words: ``where are they?'' \ \ Slightly unpacking this:
if there are technological civilizations\footnote{Here, I distinguish
between ``technological civilizations'', defined as those that can send
probes and/or beings to other solar systems, and
``scientific civilization'', defined as those that can generate
and detect electromagnetic signals.  Our civilization is still at the
``scientific stage'', but we can plausibly expect to reach the
``technological stage'' within a few centuries.}
in our Galaxy, why have they not contacted us?

If the question is formulated in this way, there are many potential
answers, such as (1) no interest in contacting us,
(2) conscious decision (\`a la {\it Star Trek}'s ``prime directive'') not to
interfere with other civilizations, (3) no interest in Galactic exploration,
(4) inability to travel
large distances (perhaps exacerbated by low density of planets
that can be colonized as bases).  And of course, there is the
very dark explanation, that all scientific civilizations
auto-destruct prior to colonizing other planets.

However, the real question, even within the context of Galactic
civilizations, is not why they failed to contact us, but why
they did not colonize Earth $1\,$Gyr ago.  If such civilizations
formed easily, then there were essentially as many opportunities
to do so around stars that formed a Gyr before the Sun as from those
having similar ages to the Sun because the rate of star formation,
with similar metal abundance, was about the same 5.5 Gyr ago as 4.5 Gyr ago.
Then, reasons (1) and (2) become irrelevant.  That is, they would
have arrived at Earth, found an excellent colonization spot that was
inhabited only by single-cell organisms, and transformed the planet
according to their own needs.  When FP is reformulated
in this way, there are really only four answers:  technological
civilizations are very rare (zero or a few per galaxy); planetary
colonization is precluded by some physical factors (such as
low density of colonizable planets or physical laws that effectively
prohibit inter-planetary travel); universal lack of interest in
colonization; or all scientific civilizations auto-destruct prior to,
or in the course of becoming, technological.

The last two reasons require universal outcomes from completely
independent social phenomena.  For this reason they are intrinsically
implausible\footnote{There are other suggested answers to FP that
also require universal outcomes from independent social phenomena, such
as the ``dark forest'' hypothesis \citep{darkforest},
in which all technological civilizations
try to hide themselves to avoid alien attack.  These are likewise plausible
only if the number of such civilizations is small.  In addition, regarding this
particular hypothesis, if the MW is thoroughly colonized (rather
than investigated by electromagnetic signals), then there is 
nowhere to hide.}.  Moreover, I will argue specifically 
against ``universal lack of interest'' in Section~\ref{sec:imperative}.
This leaves only two reasons: very few technological civilizations
per galaxy or extreme difficulty of colonization.  Note that the first
of these two is strongly coupled to the rejection of arguments
that require universal outcomes of social development.  That is,
if there were only two or three technological civilizations per galaxy,
then it is easy to imagine that they all auto-destructed or decided
not to colonize.  This becomes much less plausible when there are
dozens, keeping in mind that it only takes one technological
civilization to colonize an entire galaxy.

The problems posed by FP become much more severe,
once it is recognized that inter-galactic colonization at $0.1\,c$
is feasible.  Of course, in this paper, I have only shown
that this is feasible from the standpoint of energy requirements.
To actually colonize would require reconstruction of beings (humans,
trans-humans, or androids) from digital data.  However, given the
progress in this direction in the last 50 years, it is hard to
believe that it will not be achieved in the next kyr, or
at least the next Myr, which is still very short compared to the
several Gyr over which stars are forming.  There are about 500
MW-like galaxies within 30 Mpc, i.e., within 1 Gyr travel
time at $0.1\,c$.  Hence, if there were even one technological
civilization per galaxy (on average) then there would be 500
such civilizations in this zone, and perhaps 150 that would have
had the time needed to colonize Earth a Gyr or more before the present.

From this we can conclude that the number of technological civilizations
per galaxy is small compared
to one, or else reconstruction of beings from digital data is not
physically possible.

While this discussion has been couched in developments taking place
over Myr and Gyr, it actually has immediate practical implications:
searching for scientific civilizations in our immediate neighborhood
(e.g., 1 kpc) is very unlikely to succeed.  The chance of finding one is
less than $(1\,\kpc/R_0)^2/100\sim {\cal O}(10^{-4})$.  Rather, interplanetary
exploration should focus on sending probes to other planets to search
for living organisms, to either be analyzed locally or brought back
in sample-return missions.  The probability of success is orders of
magnitude higher, and the scientific, economic, and cultural
implications of such discoveries would be staggering.  Serious effort
in this direction would require investments of many trillions,
and many generations would pass before real results were obtained.  But these
timescales are minuscule compared to the history of humanity, and even
smaller compared to its future.  When results eventually start to pour in,
today's childish fantasies about ``Contact'' with alien intelligences will
be seen as an amusing footnote to the birth pangs of real exploration.

{\section{{The Intergalactic Imperative}
\label{sec:imperative}}

It may become feasible to launch probes toward M31 within a few
hundred years, but there will be no compelling reason to do so at that
time.  For one thing, the first leg of the voyage (i.e., Earth to
SgrA*) will take a few Myr, so that waiting even an additional century
would likely yield technological advances that could shorten this part
of the voyage by many (or hundreds) of millenia.  Moreover, while such
exploration could be motivated purely by intellectual curiosity, most
exploration in the past has been driven by multiple considerations,
not all so uplifting.

A realistic assessment of the urgency of M31 exploration and colonization
will require a detailed inventory of life forms in our own Galaxy.
Making use of a system of WD-binary ``slingshot hubs'', it will
be possible to explore the MW at $0.002\,c$ and so to survey the 
biological content of the entire MW in $\la  50\,$Myr.  Depending
on the richness of that content, substantial progress might be made
a factor of 10 sooner (based on the 1\% of the Galaxy that
is 10 times closer).  This survey will permit direct evaluation
of most terms in the Drake equation.  For example, we might find
that, apart from Earth, the MW is entirely sterile.  Or, we might
find that unicellular life is ubiquitous, but there are only a handful
of planets that have reached the stage of an energy-rich environment
(probably based on free oxygen, but maybe something else), and none
have multi-cellular animals.  Such survey outcomes would support
the hypothesis that technological civilizations are truly rare, and
plausibly non-existent within 30 Mpc (or whatever distance could
plausibly be traveled in a Gyr based on technologies developed
over the next few Myr).  Under these conditions, there would
be no compelling necessity to colonize of M31, and efforts might
well be focused on the prodigious opportunities for colonization
in the MW.

However, it is also possible that the survey will find thousands or
millions of planets with multi-cellular organisms, with progressively
smaller subsets reaching progressively higher levels of development.
This would permit not only direct measurement of many known terms
in the Drake equation, but also the identification of new ones.
For example, perhaps there will be dozens of planets with complex
agricultural civilizations that have existed for many Myr,
but with no mathematical physics, so no
concept of Maxwell's equations, let alone quantum mechanics.  Perhaps,
instead (or in addition), there will be dozens of planets with the buried
ruins of complex civilizations that had reached (or approached)
the technological stage.

Contrary to current infantile thinking, either of these possibilities
would be extremely alarming.  They would mean that the raw material
for technological civilizations is extremely widespread and only one
or two steps from those beings gaining the ability 
to leave their planets and start colonizing.
In particular, the latter possibility (dozens of auto-destructed civilizations)
would present a lurid picture of what such civilizations would be
capable of if they ever did start colonizing.

One might hope that any alien technological civilizations that were
encountered would be populated by angelic beings (or trans-beings),
interested only in helping anyone they met to achieve a higher level
of consciousness.  Indeed, this might turn out to be true.  However, without
any specific evidence in favor of this pleasant thought, one should
prepare for the opposite possibility, that their inter-galactic
colonization was at least driven by defensive military considerations
(see below), if not a direct outcome of their victory in their own
internal wars and/or inter-alien wars.

The only way to block such alien colonization efforts in the MW would
be to occupy all planets where the aliens could potentially gain a foothold.
And then to begin occupying neighboring galaxies, both to forestall,
and to warn of, their approach.  The fact that the characteristic
timescale of such an approach is Gyr (i.e., the timescale of
galaxy evolution), would not lessen the urgency because the timescale
of colonizing our own ``sphere'' of galaxies within 30 Mpc 
would be essentially the same.  Note that the very act of setting
up such a defensive colonization system would likely be seen
as potentially hostile by alien civilizations, and 
the anticipation of just such possibilities would have been
one (of possibly several) of their own motivations for expanding.

{\section{{BH Slingshot: A Poor Man's Wormhole}
\label{sec:wormhole}}

About 30 years ago, I attended a public lecture about Einstein
by a famous mathematical physicist.  To illustrate 
the Principle of Equivalence, the speaker invited the
audience to consider a trip in an ``elevator cabin'' that was
falling freely through an evacuated tunnel going straight through
Earth (presumably on the polar axis, though he did not specify).
If the cabin had no windows, the speaker claimed that people
in the elevator would have no way to tell that they were accelerating
and decelerating through Earth from one side to the other, 
and that they were not floating freely in space.

After the lecture, I approached the speaker and politely told him
that there was such a method to distinguish the two situations: simply
hold out ones arm and release a ball.  For the person falling through
Earth, the ball would fall toward his feet and then rise again
to his arm\footnote{If Earth were of uniform density, this motion
would be exactly synchronized with the position of the elevator relative
to Earth's center.}, while in free space, the ball would just stay
by his arm.

The speaker insisted that he was correct, and, perhaps mistaking
me for a member of the general public, pedantically lectured me
for about a minute that there were no known exceptions to
the Principle of Equivalence.

In fact, the elevator example does not contradict the Principle
of Equivalence because this principle applies only to {\it local}
observers, where ``local'' means local in 4-space, i.e., in
both space and time.  ``Local in space'' means small compared to the
gradient of the gravitational field, and ``local in time'' means
short compared to a dynamical time.

In particular, tidal effects will always generate order-unity
phenomena if allowed to persist unopposed for a dynamical time.

At first sight, the BH slingshot appears to be a ``cost free''
method for accelerating to $0.1\,c$ in of order 0.1 seconds.
The spacecraft is in free fall and so does not experience any
inertial forces.  Hence, it might appear that there could
be a human passenger who likewise is brought to $0.1\,c$
in 0.1 seconds (without ``feeling'' any accelerating forces), 
compared to the 36 days that this would require
at $\sim 1\,g$ acceleration.  In starting out with this
naive reasoning, one can almost hear Milton Friedman
cautioning that ``there is no such thing as a free lunch'',
but within the Equivalence-Principle elevator 
framework of the above-mentioned speaker, one is tempted
to ignore tides.  In reality, however, the BH slingshot operates
(by construction) on a free-fall timescale, so tides have an order
unity effect.  These order-unity tidal effects are 
Friedman's ``cost'' of the ``free''
acceleration to semi-relativistic speeds.

A (say, spherical) collection of particles with no forces
between them will, in slingshot  in-fall, change their collective
shape into an elongated ellipsoid, and then splatter during exit.
This is true in exactly the same way for a BH slingshot as a 
Jupiter slingshot.  The difference is only in the level of inter-particle
forces that are required to counter the tidal forces.  For a Jupiter
slingshot, the roughly spherical human brain would at all times
maintain its shape in the face of the tidal forces, whereas for
a BH slingshot it would not (even if the skull could be kept in
place by mechanical support).

In this sense, the BH slingshot can be considered intermediate between
a Jupiter slingshot and wormhole transport in efforts to evade the
normal requirements of transport by ``$F=ma$'' via more complex
gravitational fields.  There is always a ``price'' (insurmountable
for material objects in the case of wormholes), with the benefit
generally proportional to the cost.

However, the BH slingshot method could be made 
more human friendly by considering
larger orbiting BHs.  I have assumed that the BHs orbiting
SgrA* have $M=5\,M_\odot$, but recent work has shown that the BH
mass function may well be centered at twice this value and cover
a broad range of masses.  And in the dense environment of the
Galactic center, there could be substantially more massive BHs,
e.g., $M=30\,M_\odot$, such as have been detected in LIGO
observations.  At fixed infall velocity (so fixed $M/q_{\rm bh}$), the tidal
forces scale as $M/q_{\rm bh}^3\propto M^{-2}$, so they would be $(30/5)^2=36$
times weaker than I have calculated.  This could be used to increase
the ejection speed by a factor $6^{1/3} = 1.8$
(see Equation~(\ref{eqn:neval})), or it could be used
to have 36 times less stressful tides.  It is unlikely that this
would reach a level that was safe for humans, and in any case, there
does not (at present) seem to be any application for
such human slingshots at the Galactic center.  Nevertheless,
a less stressful slingshot could be important for transporting
more complex equipment.

The idea that tides produce order-unity effects if allowed to
develop over a dynamical time (and so ``violate'' a misconstrued
version of the Principle of Equivalence) seems obvious.  However,
many ``obvious'' things remain unrecognized until they have
been pointed out (or, as in the case of the speaker,
even after they have been pointed out).  More to the point,
the significance of ``obvious'' ideas is often missed until
they are explicitly formulated.

Stating this ``obvious'' point in another way, gravitational
phenomena that arise only on a dynamical timescale are tidal in
nature and are therefore not subject to the Principle of Equivalence.
An interesting example is the emission of \citet{hawking75} quanta. 
The BH temperature (in units $\hbar=c=1$) is $T= 1/8\pi M$, so
(including those quanta that are eventually reflected back into the BH),
the number of quanta emitted per dynamical time 
$t_{\rm dyn} \simeq \sqrt{R_{\rm Sch}^3/M} = \sqrt{8}M$ is
\begin{equation}
{\cal N} = {6\zeta(3)\over \pi}T^3 4\pi R_{\rm Sch}^2\sqrt{8}M
= {3\sqrt{2}\zeta(3)\over 8\pi^3} \simeq 0.02 .
\label{eqn:dyn}
\end{equation}

{\section{{Conclusions}
\label{sec:conclude}}

Intergalactic exploration by robotic spacecraft at speeds of $\sim 0.1\,c$
is feasible based on technologies that either exist today or can be
developed based on our present understanding of physics and materials.
The key intellectual innovations that allow this are first, using the
small BHs that orbit the supermassive BHs at the centers of galaxies
as sling shots, and second, using gravitational microlensing to navigate
these sling-shot encounters to high ($\sim 1\,$km) precision.

Conceptually similar, but much weaker sling shots from WD binaries
could enable exploration of the MW at $\sim 0.002\,c$, permitting
complete exploration of our Galaxy in about 50 Myr.

I have argued through several steps that the practical possibility
of intergalactic travel implies that either ``technological civilizations''
are rare ($\la 0.01$ per MW-like galaxy), or it is physically impossible
to reconstruct intelligent beings (human-like, trans-human-like, or robotic)
from digital data.  Here a ``technological civilization'' is defined as
one that is capable of sending robotic spacecraft to other solar systems.

To briefly recapitulate, if there were an average of even 1 such
civilization per galaxy, then there would be 150 capable of reaching Earth
at a time when Earth had only unicellular life.  Hence, there would be
no compelling reason not to colonize it.  If astro-biological
investigation of their own galaxy showed that technological civilizations
were feasible (even if not actually present there), these aliens would face
an ``intergalactic imperative'' to colonize all planets in whatever
galaxies they could reach (including ours) as a defensive measure.
Always provided that it is actually possible to reconstruct beings from
digital data, this would be an exponential process, limited only by
spacecraft travel time, using essentially
unlimited resources of other planets, so there would be no physical
barriers to stop this process.  All counter-arguments rest on
theories of universal social development by completely independent societies.
Our own experience on Earth, even with physically very similar people
in different places, shows that such convergent social evolution is
beyond unlikely.

The main immediate conclusion is that looking for electromagnetic
signals from alien civilizations in the solar neighborhood is not likely
to be productive.  While such ``scientific civilizations''
(i.e., like our own) are a few steps below ``technological civilizations''
and therefore somewhat more common, this effect is actually small.
The time between the two phases is likely measured in centuries,
compared to the Gyr timescales of galactic evolution that govern
the ``intergalactic imperative''.  Only if the overwhelming majority of
all such civilizations auto-destructed would there be any serious
chance of finding one in our own neighborhood.

Hence, the main conclusion that one could draw from the detection
of even one such ``scientific civilization'' is that we are
not long for this world.
  
\acknowledgments

I thank Subo Dong for stimulating discussions.




%

\begin{deluxetable}{llll} 
\tablecolumns{6} 
\tablewidth{0pc}                                                      
\tablecaption{\textsc {M31 Voyage/Colonization}} 
\tablehead{
\colhead{Phase} &
\colhead{Dist.} & 
\colhead{Speed} &
\colhead{Time}\cr
\colhead{ } &
\colhead{(kpc)} & 
\colhead{($c$)} &
\colhead{(Myr)}\cr
}                                          
\startdata
Earth--(WD$^2$)    & 0.03 & 0.0007 & 0.15 \\
(WD$^2$)--SgrA*    & 8.1  & 0.002 & 13 \\ 
SgrA*-- M31 & 780  & 0.1    & 25 \\
Explore/Colonize   & 20   & 0.002 & 33 \\
M31--Earth$^{\rm a}$& 780  & 0.1    & 25 \\
\enddata                                                              
\tablecomments{a: optional}
\label{tab:tab1} 
\end{deluxetable}

\begin{figure}
\plotone{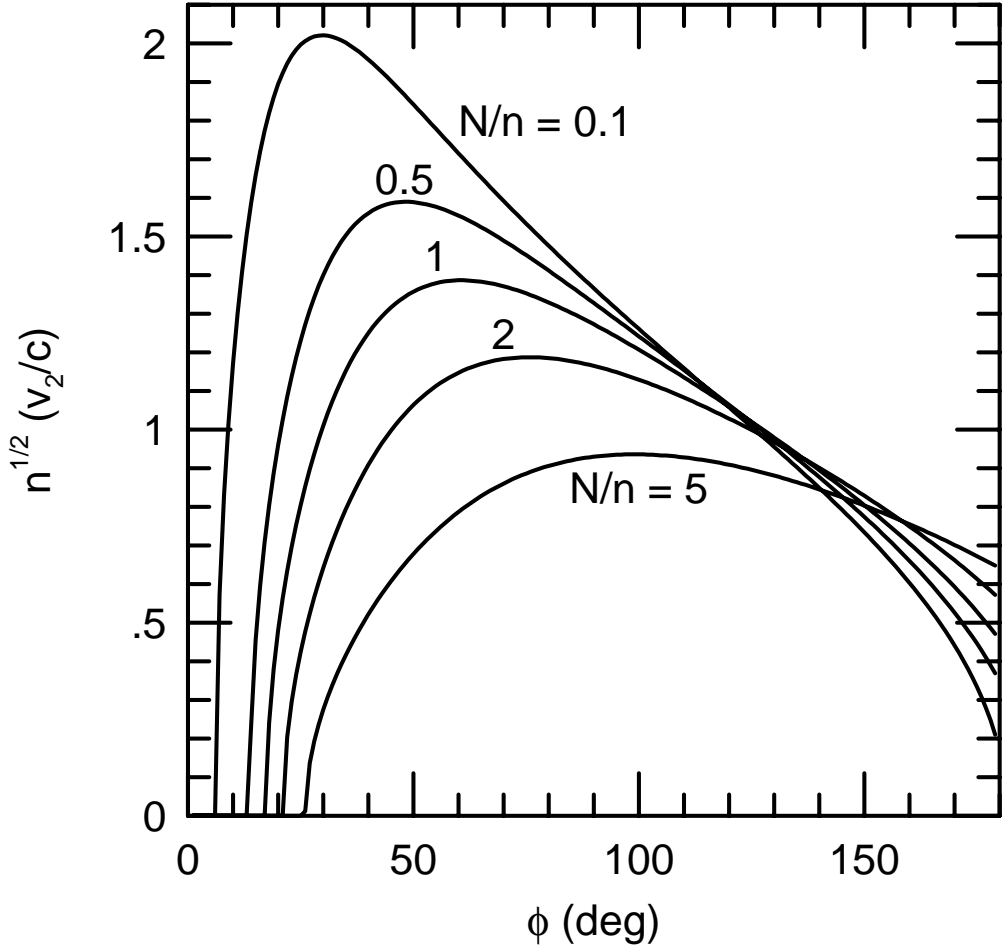}
\caption{Ejection speed of test particle, $v_2$, scaled to speed of the
spacecraft at peribothron of the small BH ($c/\sqrt{n}$) as a function of
the angle $\phi$ of the spacecraft velocity relative to the small-BH
orbital direction.  Curves are shown
for five values $N/n$, where $n$ is the number of small-BH gravitational radii
of the test particle at peribothron and $N$ is the number of large-BH gravitational radii of the orbit of the small BH.  As shown in Section~\ref{sec:tidal},
$n$ is limited by tidal effects and the tensile strength of the spacecraft,
but could plausibly be $n\sim 100$.  The curves show that ejection
velocities $v_2\sim c/\sqrt{n}$ are generally achievable, provided that
$N\la n$.
}
\label{fig:v2}
\end{figure}


\end{document}